%% file: paper.tex
\pdfoutput=1
\documentclass[sigconf,compact]{acmart}
\usepackage{
    graphicx,
	url,
    hyperref,
    cleveref,
    adjustbox,
    float,
    multirow
}
\def\BibTeX{{\rm B\kern-.05em{\sc i\kern-.025em b}\kern-.08emT\kern-.1667em\lower.7ex\hbox{E}\kern-.125emX}}
    
\copyrightyear{2019}
\acmYear{2019}
\setcopyright{ccby}
\acmConference[CARRV 2019]{Third Workshop on Computer Architecture Research with RISC-V}{June 22, 2019}{Phoenix, AZ}
\acmDOI{}
\acmISBN{}

\newcommand{\sfence}{\texttt{SFENCE.VMA}}
\newcommand{\newcsr}{MASI}

\begin{document}

\title{Fast TLB Simulation for RISC-V Systems}

\author{Xuan Guo}
\email{Gary.Guo@cl.cam.ac.uk}
\affiliation{%
  \institution{University of Cambridge}
  \city{Cambridge}
  \country{UK}
}

\author{Robert Mullins}
\email{Robert.Mullins@cl.cam.ac.uk}
\affiliation{%
  \institution{University of Cambridge}
  \city{Cambridge}
  \country{UK}
}

\renewcommand{\shortauthors}{Guo, et al.}

%
\begin{abstract}
\input{sections/abstract}
\end{abstract}

%
%

%

\maketitle

\input{sections/intro}
\input{sections/related}
\input{sections/implementation}

\input{sections/performance}
\input{sections/application}
\input{sections/shared}
\input{sections/conclusion}

%
\bibliographystyle{ACM-Reference-Format}
\bibliography{paper}

\end{document}

%% file: sections/abstract.tex
Address translation and protection play important roles in today's processors, supporting multiprocessing and enforcing security. Historically, the design of the address translation mechanisms has been closely tied to the instruction set. In contrast, RISC-V defines its privileged specification in a way that permits a variety of designs.

An important part of the design space is the organisation of Translation Lookaside Buffers (TLBs).
This paper presents our recent work on simulating TLB behaviours in multi-core RISC-V systems\footnote{Available at https://github.com/nbdd0121/TLBSim under BSD-2 licence.}.
Our TLB simulation framework allows rapid, flexible and versatile prototyping of various hardware TLB design choices, and enables validation, profiling and benchmarking of software running on RISC-V systems.
We show how this framework can be integrated with the dynamic binary translated emulator QEMU to perform online simulation. When simulating complicated multi-level shared TLB designs, the framework runs at around 400 million instructions per second (MIPS) when simulating an 8-core system. The performance overhead compared to unmodified QEMU is only 18\% when the benchmark's L1 TLB miss rate is 1\%.

We also demonstrate how this tool can be used to explore the instruction-set level design space. We test a shared last-level TLB design that is not currently permitted by the RISC-V's privileged specification. We then propose an extension to RISC-V's virtual memory system design based on these experimental results.

%% file: sections/intro.tex
\section{Introduction}

RISC-V is an simple, extensible, general-purpose and open instruction set architecture (ISA). 
In the past year, RISC-V support has been upstreamed in many open-source software projects, such as the Linux kernel, GNU toolchains, LLVM, etc. We also see an emerging number of open-source RISC-V processsors debuted in the past year, such as Ariane from ETH Zurich, and SweRV from Western Digital Corporation. There are also a number of open-source system-on-chip (SoC) platforms being actively developed, such as lowRISC from lowRISC CIC.

Most of the existing open-source RISC-V implementations are simple cores which focus more on the microcontroller use case, such as RI5CY from ETH Zurich. However, there is also a demand for high-performance, Linux-capable RISC-V SoC implementations for personal computing or even server workloads. Address translation and protection, or virtual memory addressing (VMA) support, is one of the core components of such a system.  Processors supporting virtual memory often employ translation caches such as Translation Lookaside Buffers (TLBs) to cache virtual-to-physical mappings, avoiding the cost of high-latency accesses to in-memory address translation data structures. TLBs are crucial to system performance due to the penalties associated with misses \cite{barr2010translation}\cite{basu2013efficient}.
We investigated the few existing open-source implementations with supervisor mode and discover that their VMA and TLB support is relatively simple. They use a single-level fully-associative TLB. Furthermore, the same TLB design is duplicated for instruction and data accesses.
In addition, the current version of Linux kernel (5.0) only uses a limited subset of VMA features that RISC-V provides.
This work aims to help validate and profile different TLB design decisions in simulation using real workloads, generating results to guide software, hardware and ISA design decisions. This will be particularly important for the high-performance many-core RISC-V systems that we anticipate will soon be produced. 

In this paper, we present a framework for simulating TLB behaviours, and conduct experiments to explore the design-space of RISC-V virtual memory systems. We make the following contributions:
\begin{itemize}
    \item We demonstrate how QEMU can be modified to perform fast TLB simulation with only a minor performance loss. Even when simulating complex multi-level TLB designs, we incur only a 18\% performance overhead compared to vanilla QEMU, and can achieve around 400 MIPS when simulating an 8-core system.
    \item We use the tool to log and categorise the Linux kernel's use of \sfence{}, RISC-V's TLB flush instruction. We suggest that an extra constraint is placed upon the hardware to aid more efficient software implementations.
    \item We investigate the case for shared TLBs and a global ASID space. We find that shared TLBs have a concrete advantage over private TLBs, and that a global ASID space can increase gains further. We then propose an extension to permit the use of such a design for RISC-V.
\end{itemize}

%% file: sections/related.tex
\section{Background}
\label{sec:related}

\subsection{VMA in RISC-V}
\label{sec:riscv-vma}
The privileged specification 1.10 \cite{waterman2016risc} of RISC-V defines the address translation structure of in-memory page tables to support VMA. Depending on machine word-length and configuration, virtual addresses are translated to physical addresses using two, three or four levels of page table. A control and status register (CSR) called SATP (Supervisor Address Translation and Protection) is used to store the base address of the root page table. In addition to the base address, a paging mode field and a 16-bit address space identifier (ASID) field (9-bit in 32-bit RISC-V) are packed into the SATP CSR allowing these fields to be changed atomically. The ASID field can be used to distinguish between TLB entries to avoid the need to flush TLBs during context switches.

RISC-V also includes a \sfence{} instruction to flush TLBs. \sfence{} takes two optional register operands to specify the ASID and virtual address to flush. Implementations may choose to ignore register operands and always perform a full TLB flush.

Unlike almost all other architectures which describe similar instructions as flushing translation cache entries from TLBs, RISC-V defines the instruction as a fencing instruction after which the previous modifications to address translation data structures are required to be honoured by the hardware. The specification is worded in such way deliberately to allow a wide design space. However, as a result of such definition, an implementation can technically cache invalid entries in its TLB while still being RISC-V compliant.
We will revisit this issue in Section \ref{sec:soft_prof} and explore its implications using our TLB simulator.

\subsection{ASID Space}
\label[section]{sec:asid}
RISC-V, in its current state, also mandates that all hardware threads (harts) need to have their own private ASID spaces.
The concept of hart-local ASID spaces is frequently used in other ISAs which add ASID support as a later extension, e.g. x86 \cite{intelmanual}. Without ASID support initially, each core naturally has separate address translation units. For backward compatibility considerations, ASIDs are kept local to each core when they are added.
On the other hand, for architectures that are designed to have ASID support from day one, such as ARMv8, the ASID space is global to all processors, and remote TLB cache shootdown is also supported.

RISC-V's privileged specification has undergone some large scale changes, and there is already software that makes use of its virtual memory system without having ASID support.
As RISC-V requires CSR fields unknown to software to be filled with 0, it happens that software naturally uses ASID 0 when ASIDs are introduced.
The backward compatibility burden essentially requires ASID 0 to be treated as local to each hart.
Following our suggestions about a global ASID space, the committee has decided to place a forward-compatibility requirement on software so that non-zero ASIDs should have consistent meanings across different harts, and software should not use ASID 0 if it decides to use other ASIDs.

\subsection{Shared TLB}
\label{sec:rel_shared}
A global ASID space allows more than remote TLB shootdown. It allows harts to share last-level TLBs, and also possibly inter-core cooperative prefetching \cite{bhattacharjee2010inter}\cite{lustig2013tlb}. Bhattacharjee, et al. investigated the potential performance gains of using a single shared L2 TLB when compared to multiple per-core private L2 TLBs (with the same number of total entries when aggregated) \cite{bhattacharjee2011shared}. Their result showed that shared TLBs can consistently outperform private TLBs, and on average the L2 miss rate is reduced by 27\%. Li, et al. reported that PS-TLB can achieve similar performance \cite{li2013ps}.

A shared TLB can be most efficiently implemented if the ASID space is shared globally. As mentioned in the previous subsection, the global ASID space is not currently permitted by RISC-V. We therefore also show how a shared TLB can be designed while being compliant with the current architecture specification in Section \ref{sec:shared}, and use our framework to simulate the performance between different designs. The experimental results are then used to propose an extension.

%% file: sections/implementation.tex
\section{Simulator Implementation}

\begin{figure*}[ht]
    \includegraphics[width=0.8\linewidth]{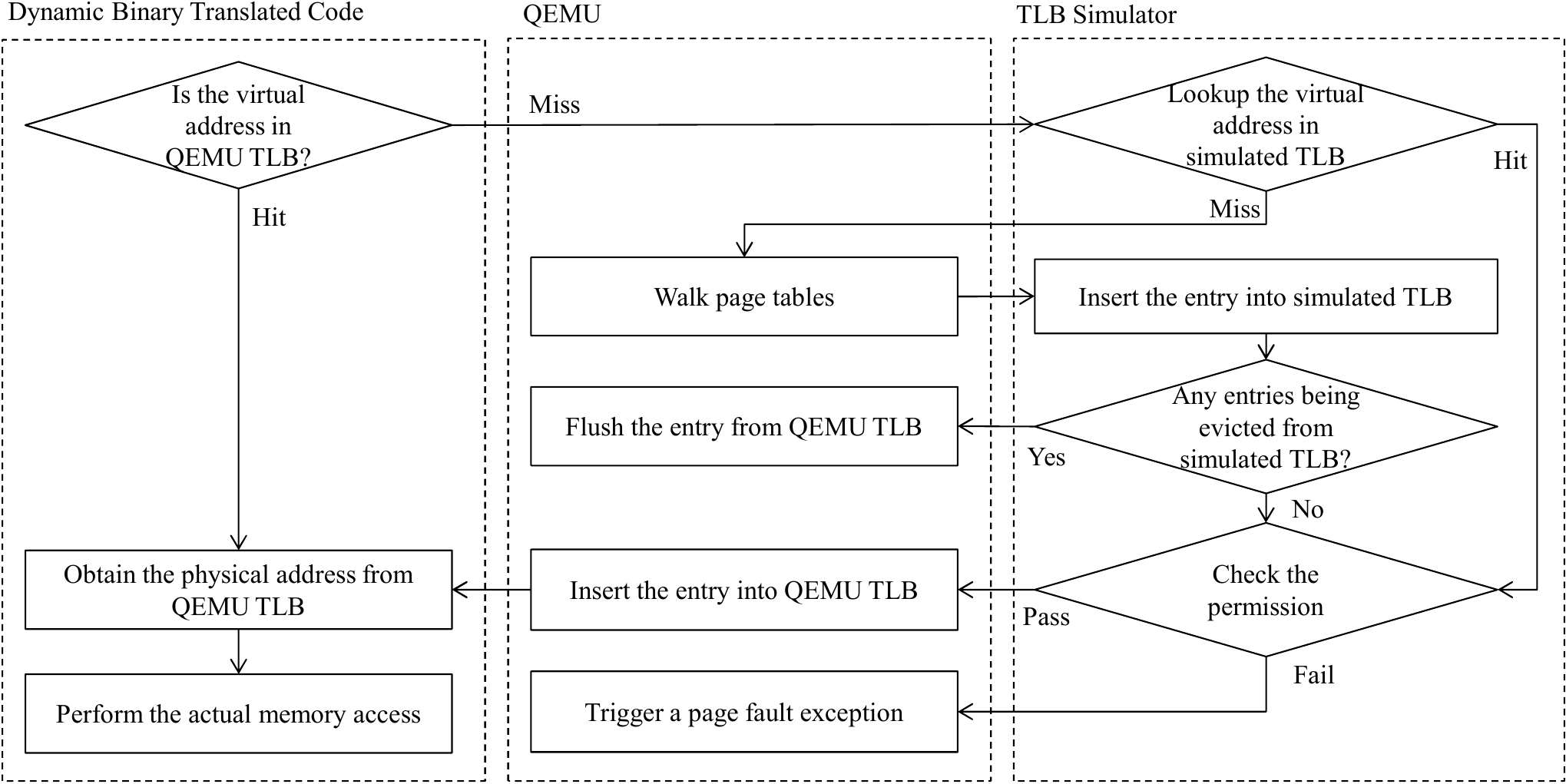}
    \caption{Control flow of a guest data memory access, with only L1 caches simulated}
    \label[figure]{fig:flowchart}
\end{figure*}

\subsection{Inclusive L1 TLB}

When running code of a non-native ISA, QEMU uses its tiny code generator (TCG) to perform dynamic binary translation (DBT), translating guest binaries to host binaries. When TCG is compiled with soft MMU support (i.e. doing full system emulation), it uses a technique similar to hardware direct-mapped TLBs to accelerate code with virtual memory access. Any guest memory access instructions are translated to a sequence of host instructions which accesses QEMU's TLB-like structure first to acquire the physical address before performing the actual access. When the TLB access hits, the control flow does not leave generated code, and QEMU's helper code is invoked when the corresponding page does not exist in the TLB. It is worth mentioning that QEMU's direct-mapped TLB is only an acceleration mechanism, and it does not simulate or intend to simulate the actual address translation mechanisms employed in actual hardware.
Other ISA simulators, e.g. Spike, use similar approach to accelerate memory access when soft MMU is used.

We would like our simulation framework to be used without extensive modification in ISA simulators.
The case is especially true for QEMU, as modifying the TLB used in TCG requires a range of components to be changed.
Moreover, ISA simulators almost always use direct-mapped TLBs for performance concerns.
Traversing through associative caches, which most hardware TLB designs use, is clearly not acceptable on the hot path.

Our TLB simulation framework is therefore built as a shared library and dynamically linked into ISA simulators.
We made the design decision that our code should only be executed on the slow path.
When a memory location is accessed, the ISA simulator will query its own TLBs first (referred later as L0 TLB from our simulation's point-of-view). If L0 TLB hits, the ISA simulator can serve the request directly.
The API provided by our TLB simulation framework is called when L0 TLB misses, replacing ISA simulator's own page walking routine.

To ensure all TLB misses can be accurately recorded and simulated, we need to avoid scenarios where a memory access hits L0 TLB but would miss in our TLB.
To address this potential issue, we enforce that all entries in L0 TLB must also be included in our simulated L1 TLB.
If any entries are evicted from our TLB, we will use the L0 TLB invalidation callback provided by the ISA simulator to remove the entry from L0 TLB as well.
Figure \ref{fig:flowchart} shows the control flow of a guest data memory accesses when the user is QEMU, with only L1 caches simulated. The control flow is more complicated if multi-level TLBs are simulated.

Our approach keeps ISA simulator's fast-path almost untouched, therefore all the performance overhead of TLB simulation lies in the slow path.
As a result, only minor performance overhead is observed.
A drawback of this approach is that as we did not intercept all TLB accesses, we cannot implement least-recently used (LRU) policy in L1 TLBs.
This is a deliberate trade-off we made for high performance.
Other replacement policies that do not require all TLB accesses to be accounted, e.g. FIFO, random or even victim caches, can be supported.
We believe that the simulated result is still representative as there are researches suggesting that only small hit-rate gap exists between replacement policies \cite{mittal2017survey}.

\subsection{QEMU Integration}

Our design decision makes integration with QEMU simple. To be able to use our TLB simulation framework, only the following changes are made in QEMU:
\begin{itemize}
    \item Replace page walking routine with calls to TLB simulator;
    \item When \sfence{} is executed, call TLB simulator in addition to normal flush;
    \item Inserting code to the beginning of each DBT-ed basic block to update instruction retirement counters. We need two counters, total number of instructions executed and number of memory access instructions executed. By tracking these two counters, we can calculate number of L1 hits even when some of the lookups never reach our TLB simulator.

    To avoid the synchronisation cost by keeping per-CPU counters, the inserted code will only update per-CPU counters, and the counters are only aggregated when control is handed back to QEMU's main loop from DBT-ed code.
\end{itemize}

\subsection{Features}

We have implemented fully associative, set associative and direct mapped TLBs with FIFO replacement policy. Other replacement policies and victim caches can also cooperated within our framework. It is also possible to implement advanced techniques such as prefetching and CoLT \cite{pham2012colt} in our framework.

Besides the conventional TLB designs used in hardware, we also implement an ``ideal TLB''. Ideal TLB caches all translations that it sees, and never evicts entries. It is identical to a fully associative TLB with infinite number of entries. It is the basis of the Software Validation use case that we describe in \Cref{sec:soft_valid}.

Our framework is able to gather various statistics related to the virtual memory system, including number of misses, evictions and flushes associated with each level of TLB.
We also categorise page faults and executed \sfence{}s and provided separate counters for each.

Our TLB simulation framework is designed to be thread-safe to 
support multi-threaded ISA simulators, e.g. QEMU with multi-thread TCG (MTTCG) enabled, where all logical guest cores run in their own thread.
All statistics are gathered using counters implemented as atomic variables. Other critical data structures are protected by light-weight spin locks instead of mutexes.
We use fine-grained spin locks whenever possible to reduce possibility of contention.

We also support offline L2 TLB simulation.
A trace collector can be connected to L1 TLB models, and log TLB requests to a file.
The file can be replayed for L2 TLB simulation.
However, the offline simulation works well only on small benchmarks, therefore the experiments carried out in later sections used online simulation only.

%% file: sections/performance.tex
\section{Simulator Performance}

We evaluated the performance overhead of TLB simulation when used together with QEMU.
The performance impact is minor, as the only changes we have made to QEMU's fast path are the instruction retirement counters, which can also be turned off if these measurements are not necessary for a specific experiment or benchmark.

\begin{figure}[ht]
    \includegraphics[width=0.8\linewidth]{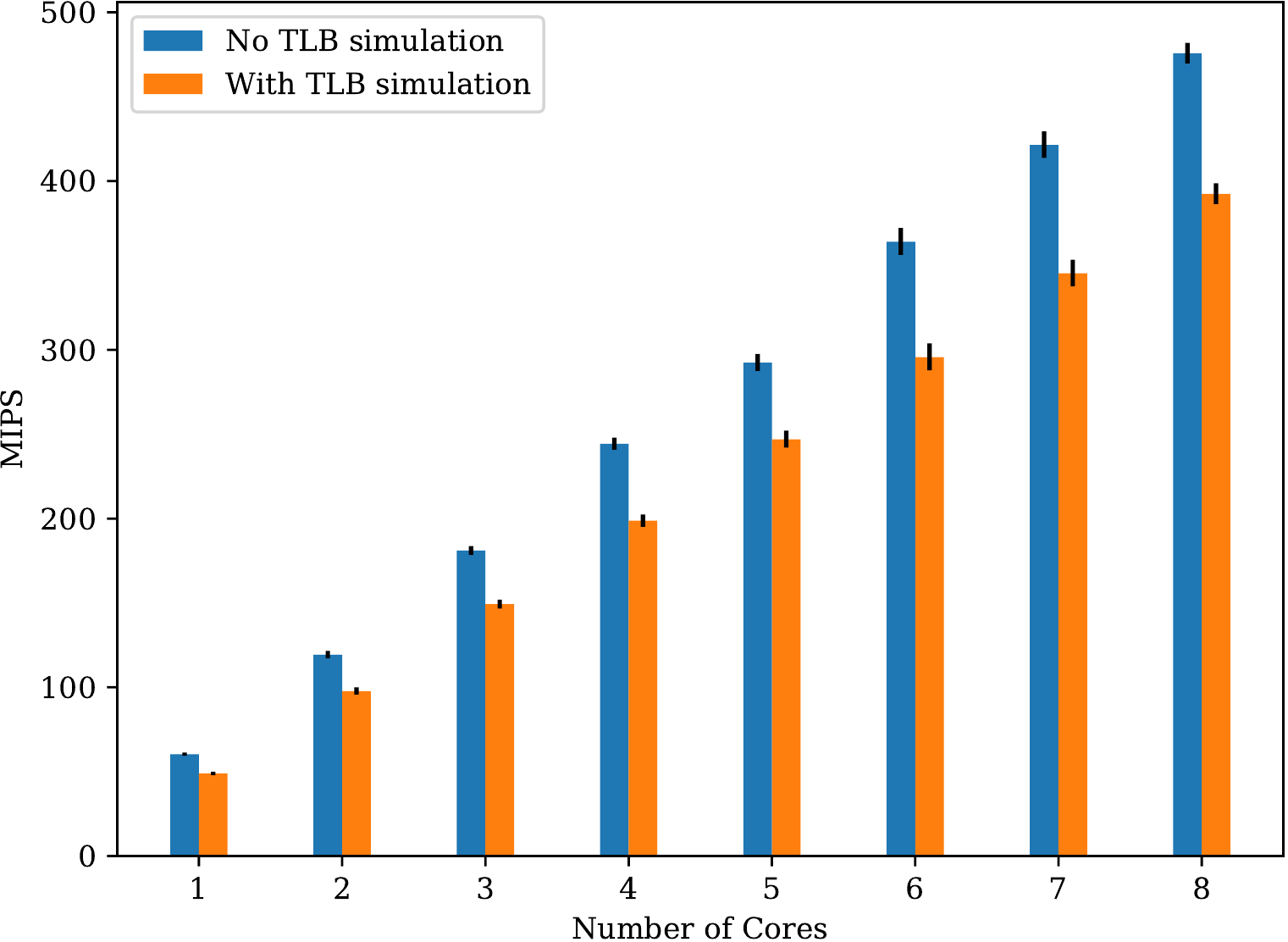}
    \caption{Performance of QEMU, with and without TLB simulation, in MIPS}
    \label[figure]{fig:mips}
\end{figure}

Figure \ref{fig:mips} shows the performance of QEMU with and without TLB simulation, when simulating a multi-level TLB design when the workload has L1 TLB miss rate of 1\%.
When running with 8 guest CPU cores, the framework achieves around 50 MIPS per core. The average performance loss compared to unmodified QEMU is 18\%.
In some experiments where L1 TLB miss rate is low, the performance loss can be as low as 2.5\%.

The performance we achieve is 2860x faster than the gem5 simulator at 175 KIPS and 125,000x faster than the Chisel C++ RTL simulator at 4 KIPS \cite{ta2018simulating}. The numbers show that if only the virtual memory system is of concern, using our TLB simulation framework provides a significant speedup.

%% file: sections/application.tex
\section{Software Validation, Profiling and Benchmarking}

\subsection{Software Validation}
\label{sec:soft_valid}

Our simulation framework can be used to verify the operating system software's behaviour on different hardware implementations.
Our framework can be tuned to disable hardware accessed/dirty bit updates, deliberately cache invalid entries, and/or generate page faults aggressively when an invalid translation entry is found in the TLB, to test if the software has handled such cases properly.
We also implemented a set of validators on top of the ideal TLB. We can detect the following common mistakes:
\begin{itemize}
    \item ASID reused without flushing;
    \item Page table entries updated with different physical address or reduced permission, but there is no \sfence{}.
    \item Multiple ASIDs used to refer to the same page table. While not strictly wrong, it often indicates an implementation error in software, and is at least a performance issue.
\end{itemize}

The experiments carried out in Section \ref{sec:shared} requires the operating system to have ASID support.
We need to implement ASID support for Linux ourselves, as its RISC-V port does not have ASID support at the time of writing.
We used the validation feature of our TLB simulator extensively to verify and debug our implementation.

\subsection{Software Profiling}
\label{sec:soft_prof}

As discussed in section \ref{sec:riscv-vma}, RISC-V compliant implementations are allowed to cache an invalid entry in its TLB, as \sfence{} is specified as an ordering instruction, so it is needed to guarantee that a write to a previously invalid leaf page table entry (PTE) is made visible.

This is contrary to what software normally assumes.
Linux, for example, assumes that TLBs cannot contain totally invalid entries.
TLB flush routines will not be called by the kernel at all in this scenario. 
The current RISC-V port of Linux (at the time of writing, 5.0) instead uses a feature that Linux provides to allow architectures to preload the MMU (\verb|update_mmu_cache|) to perform the flush instead.
This MMU preloading code is called whenever a mapping is created, for example, when a program's text segment is mapped by the \verb|execve| system call. At this time the program is not executed yet, and it is extremely unlikely that an hardware implementation will load these TLB entries speculatively. The TLB flushes have thus most likely been superfluous. 

We are interested in how RISC-V's current ISA design can impact software, as TLB flushes can be expensive operations on many micro-architectures, including but not limited to:
\begin{itemize}
    \item Micro-architectures that always perform a full TLB flush on \sfence{}.
    \item Micro-architectures with multi-level TLBs, in which case a \sfence{} requires flushes in all TLB hierarchies.
    \item Micro-architectures with virtual cache, in which case a\\\sfence{} may also require a cache flush.
\end{itemize}

\begin{table}[ht]
    \centering
    \begin{tabular}{|c|c|c|}
        \hline
        Category & Number per MInst & Percentage \\
        \hline
        Never Accessed & 5.01 & 40\% \\
        Previously Invalid & 6.78 & 54\% \\
        Previously Non-Writable & 0.59 & 5\% \\
        Necessary & 0.13 & 1\% \\
        \hline
    \end{tabular}
    \vspace{1\baselineskip}
    \caption{\sfence{}s issued by the Linux kernel}
    \vspace{-1\baselineskip}
    \label{tab:linuxflush}
\end{table}

We use our simulation framework to investigate \sfence{}s issued by the Linux kernel.
The workload chosen is the compilation of the Linux kernel, as TLB flushes happen mostly when processes are frequently spawned and terminated.
We categorise \sfence{}s issued into 4 categories, as shown in Table \ref{tab:linuxflush}:
\begin{itemize}
    \item Never accessed page: These are mappings that are newly created and never accessed. The specification allows hardware to prefetch TLB entries anywhere in the address space, but these entries will never be fetched in practice.
    \item Previously invalid page: These are mappings which were previously invalid, and are made valid by the page fault handler. This usually applies to the stack and heap where page maps are created but memory is not yet allocated. These flushes are not necessary in a design that does not cache invalid entries.
    \item Previously non-writable page: These pages are previously non-writable, and are made writable by the page fault handler. These pages are usually results of a \verb|fork| which makes page copy-on-write (CoW). These entries may have been cached in TLBs as the result of a successful read, so it is reasonable to issue \sfence{} in this scenario.
    \item Necessary flushes: These flushes are absolutely necessary, for example, as a result of \verb|mprotect| which downgrades the permissions on the page.
\end{itemize}

All cases other than (4) can be done lazily, i.e. do not flush them until they trigger exceptions. The current Linux kernel issues \sfence{}s in all cases, including (1) and (2), based on the reasoning that TLB flushes are less expensive than page faults. However, we think that the cases behind (1) and (2) are rare if not impossible to see in actual hardware implementations. As they account for 94\% of \sfence{}s, eliminating them can improve performance when flushes are expensive. We therefore suggest RISC-V privileged specification to:
\begin{itemize}
    \item Recommend against hardware from caching invalid entries;
    \item Recommend software to assume that such case is unlikely.
\end{itemize}
We do not suggest that the possibility is ruled out completely, i.e. software should still handle spurious page faults properly, therefore no additional hardware logic is needed to deal with the case where a speculatively executed memory access causes a page fault after a page table modification (even though the case never happens in a properly implemented OS).

%% file: sections/shared.tex
\section{Hardware/ISA Design Space Exploration: The Case for Shared TLB}
\label{sec:shared}

\begin{table*}[ht]
    \centering
\begin{tabular}{|l|l|l|l|}
\hline
\multicolumn{3}{|l|}{\newcsr{} mismatch}                                                                   & Cannot share; different ASID isolation domain            \\ \hline
\multirow{6}{*}{\newcsr{} match} & \multicolumn{2}{l|}{VMID mismatch}                                      & Cannot share; different VM                               \\ \cline{2-4} 
                              & \multirow{5}{*}{VMID match} & The entry has global bit set              & Can share                                                \\ \cline{3-4} 
                              &                             & Both ASIDs are non-zero and identical     & Can share                                                \\ \cline{3-4} 
                              &                             & Both ASIDs are non-zero but not identical & Cannot share                                             \\ \cline{3-4} 
                              &                             & Both ASIDs are zero                       & Cannot share; software does not support ASID             \\ \cline{3-4} 
                              &                             & One ASID is zero and another is non-zero  & Implementation-defined \\ \hline
\end{tabular}
    \vspace{1\baselineskip}
    \caption{Look-up table for whether translation caches could be shared}
    \label{tab:masidi}
    \vspace{-1\baselineskip}
\end{table*}

We discussed about the current scenario of RISC-V's ASID design in Section \ref{sec:asid}, and mentioned the benefit of shared TLBs shown by prior arts in Section \ref{sec:rel_shared}. In this section, we explore three possible L2 TLB designs and their performance in terms of hit rates.

In all three setups, all configurations other than L2 TLB are kept identical. All simulations are carried out using 8 cores, and each core has its own separate instruction L1 TLB (I-TLB) and data L1 TLB (D-TLB). All L1 TLBs are 32-entry and fully-associative. Hardware accessed/dirty bit updates are on, and no invalid entries will be cached. Non-writable entry may be cached and trigger a page fault without falling to page walker. The three setups differ by their L2 configurations only:
\begin{enumerate}
    \item Private L2 TLB: Each core has its own dedicated 128-entry, 8-way set associative L2 TLB. Both the core's I-TLB and D-TLB are connected to this L2 TLB.
    \item Shared L2 TLB, without the assumption of global ASID space: There is a unified 1024-entry, 8-way set associative TLB. Each TLB entry is tagged with a associated hart ID, in addition to an ASID and a virtual page number (VPN).
    \item Shared L2 TLB, with the assumption of global ASID space:  There is a unified 1024-entry, 8-way set associative TLB. Each TLB entry is tagged with only an ASID and a VPN. The TLB is free to serve an entry that is originated by a hart to a different hart.
\end{enumerate}
While all setups have the same number of L2 TLB entries per core, their hardware costs differ: setup (2) requires additional tag bits for hart IDs; (1) has low requirement for communications; (2) and (3) need less logic for tag checking (as they are 8-way in total instead of 8-ways per core). These three configurations are selected for their simplicity, but our framework may be modified to simulate additional sharing mechanisms, such as PS-TLB \cite{li2013ps}.

\begin{figure}[ht]
    \includegraphics[width=\linewidth]{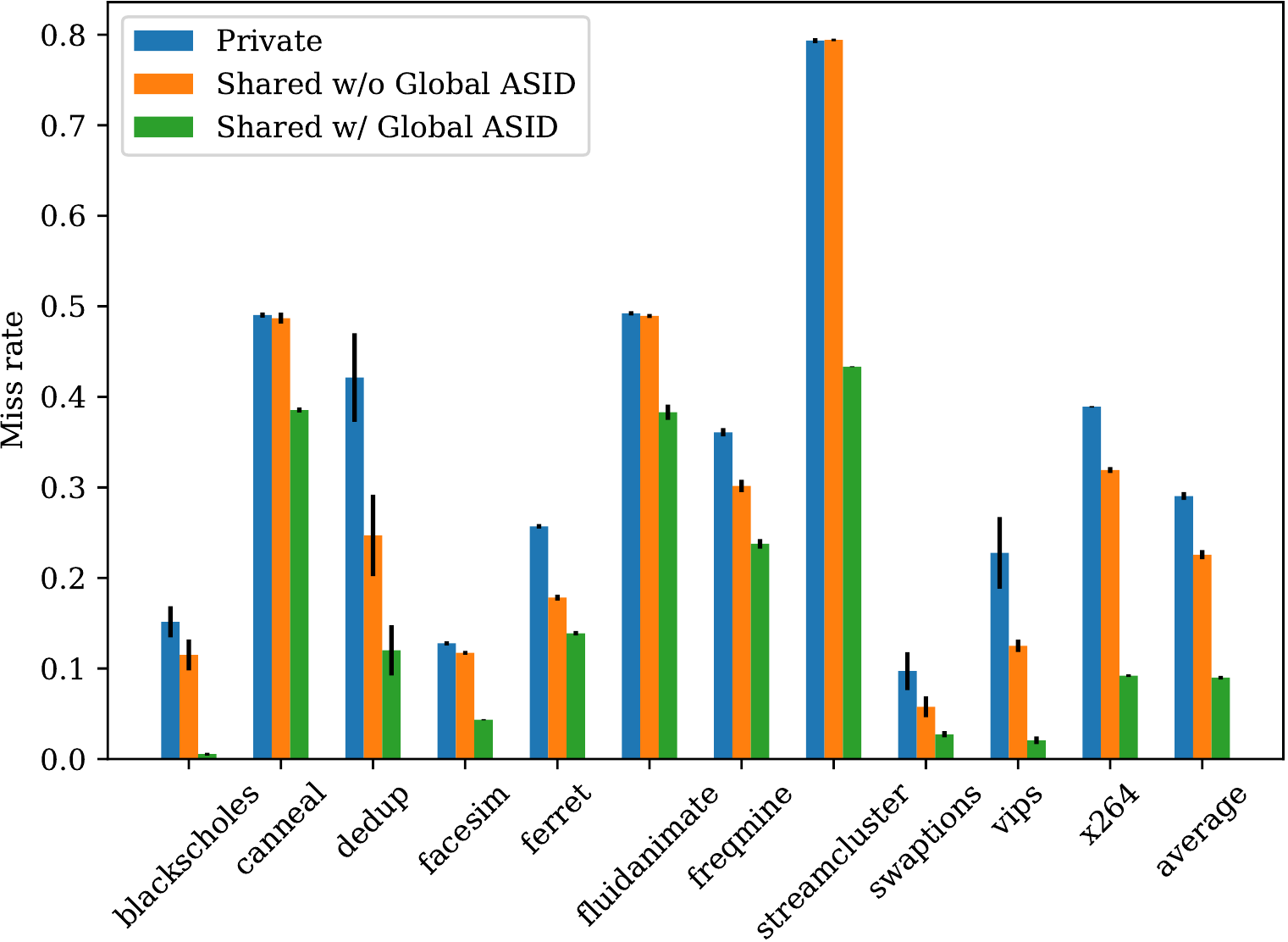}
    \caption{Local L2 TLB miss rates running the PARSEC benchmarks}
    \label[figure]{fig:parsec}
    \vspace{0.5\baselineskip}
\end{figure}

\Cref{fig:parsec} shows the results generated using the PARSEC benchmark suite \cite{bienia2008parsec} (excluding raytrace, which fails to compile) to test the scenario where memory sharing is common and frequent.
A shared L2 TLB without global ASID performs similarly or better compared to private L2 TLBs in all PARSEC benchmarks.
The average L2 miss rate drops from 29\% for private L2 TLB to 23\% for shared L2 TLB without global ASID.
We observe a major improvement in all benchmarks when shared L2 TLB is allowed to exploit a global ASID space. The miss rate drops further to 9\%, inline with our expectation. When sharing is common, global ASID space can avoid storing redundant copies, therefore can cache more entries.

\begin{figure}[ht]
    \includegraphics[width=0.8\linewidth]{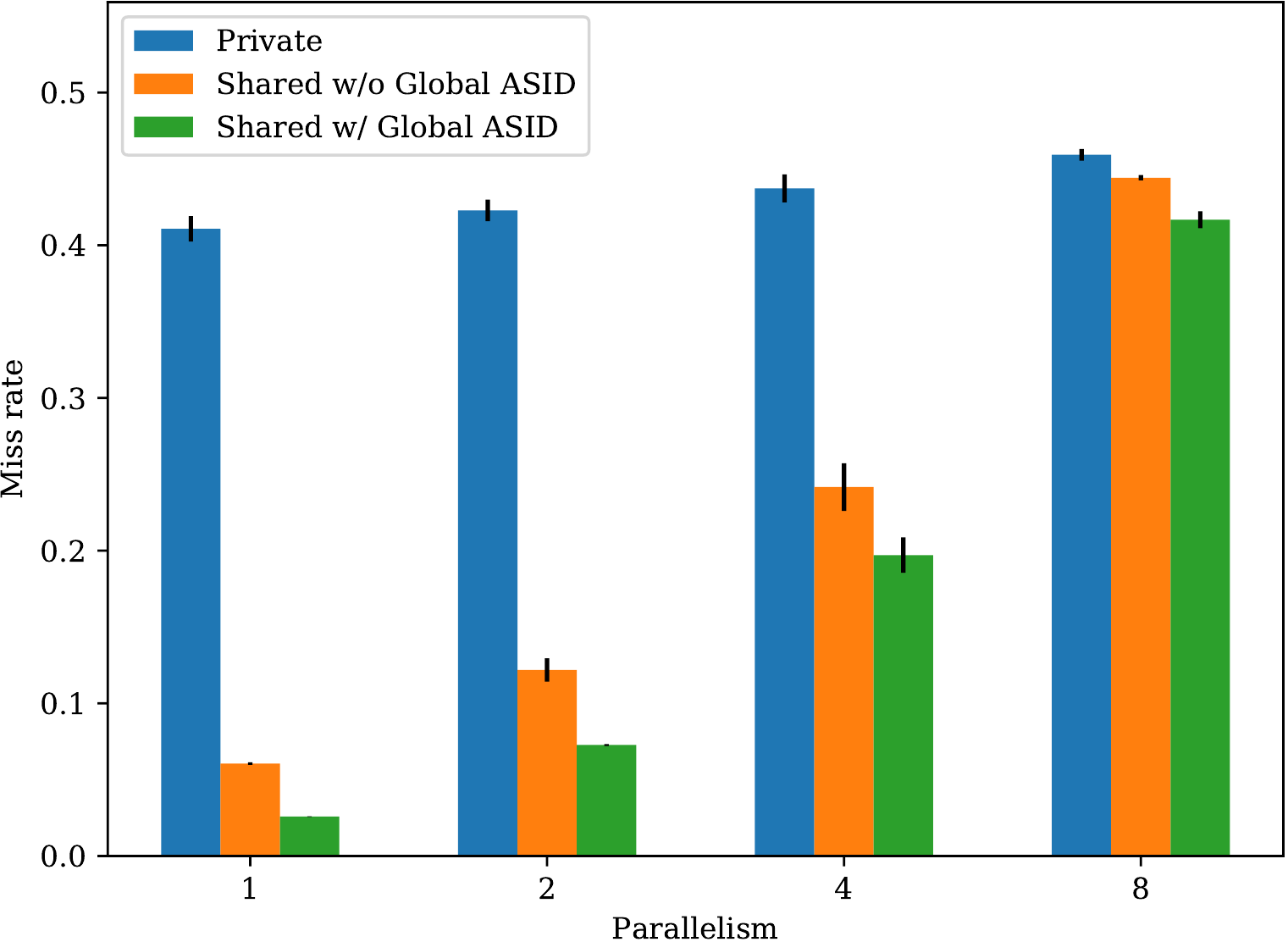}
    \caption{Local L2 TLB miss rates compiling the Linux kernel, in relation to parallelism}
    \label[figure]{fig:serial}
\end{figure}

We also use Linux kernel compilation as a benchmark.
Compiling a single C file is serial, but the whole task is embarrassingly parallel as different files can be concurrently compiled using \verb|make -j<n>|.
This benchmark tests the different TLB organisations when memory sharing and communications between cores are infrequent.
Figure \ref{fig:serial} shows the statistics we gathered.
When all 8 cores are used, the miss rate reduction is minimal as we expected.
When fewer number of jobs are used, the effective per-hart L2 TLB size for shared TLB is greater, so we see a huge performance gain over private L2 TLB. We believe that global ASID TLB outperforms shared TLB without global ASID slightly as processes are occasionally migrated between harts.
This result shows that shared TLBs can be particularly useful for low or moderate utilisation.

\begin{figure}[ht]
    \includegraphics[width=0.8\linewidth]{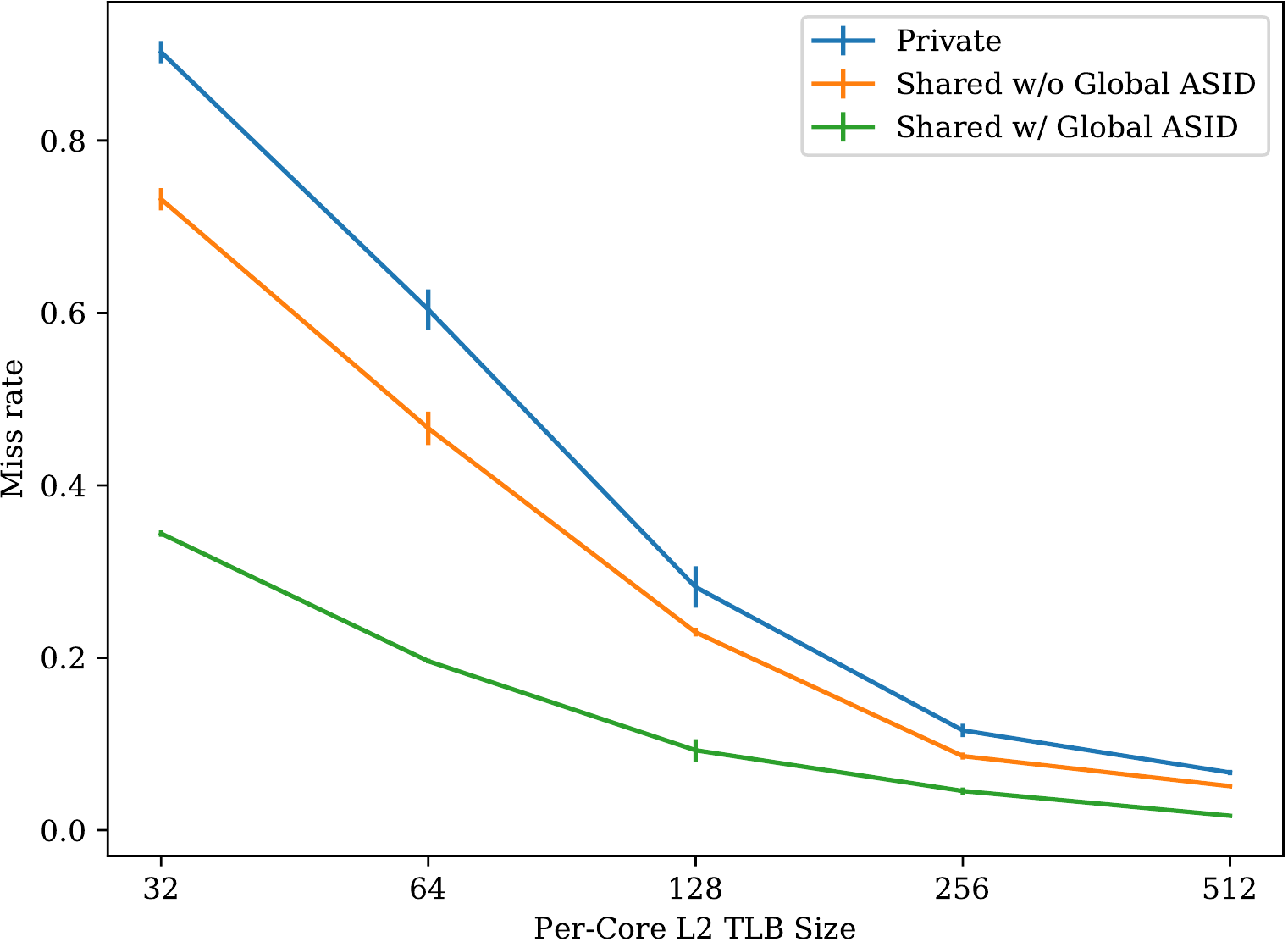}
    \caption{Local L2 TLB miss rates running the PARSEC benchmarks, in relation to per-core L2 TLB size}
    \label[figure]{fig:l2_sweep}
\end{figure}

We also explored the relationship between miss rate and per-core L2 TLB size in these three setups, with result shown in Figure \ref{fig:l2_sweep}. As we expected, the miss rate is negatively correlated to the per-core L2 TLB size in all three setups.
A more interesting observation is that the miss rate of shared TLB with global ASID is even better than the miss rate of shared TLB without global ASID with twice as many entries, and is approximately the same as the miss rate of private L2 TLB design with 4 times as many entries per core.
This suggests that, in parallel workloads, if we change a private L2 TLB design to a shared L2 TLB design with global ASID, we will be able to save $3/4$ space while preserving the miss rate.

\subsection{Extension to RISC-V}

Following empirical data from the last section, we see a clear advantage of shared TLBs with global ASID space, and believe that such design should definitely be allowed by the specification.
Global ASID space also has other potential benefits than just allowing shared translation caches.
It also permits virtual cache, which is used in some GPU designs \cite{yoon2018filtering}.
For OSes, the gain of independent hart-local ASID space has little value, and may not be worth the complexity of tracking ASID individually per hart.
If hardware supports remote TLB shootdown, by using consistent meaning of ASIDs across harts, OSes can take advantage transparently by issuing \verb|sbi_remote_sfence_vma_asid|.

To permit these designs, RISC-V's privileged specification requires a change. However, we cannot simply declare that now ASIDs are global -- this will break existing software that uses ASID 0 on all harts, and we will lose the ability to perform logic partition (LPAR) style separation on hardware without hypervisor extension support. We therefore propose an ``ASID space sharing extension'' instead:

A new CSR, \newcsr{} (Machine Address Space Isolation CSR) should be introduced, which is a write-any-read-legal (WARL) CSR register. This CSR can be implemented by all hardware implementations regardless whether they would like to use global ASID space.

The number of writable bits are implementation-defined. Non-writable bits are hardwired to constants, which may or may not be zero. Less significant bits are implemented first. Software can determine the writable part of this CSR by first writing all zeroes to \newcsr{} and read back, then writing all 1s to \newcsr{} and read back, and XORing the results. Following this procedure, the bits containing 1 in the XORed result are writable.

Table \ref{tab:masidi} shows whether two entries could be shared under the proposed extension. We deliberately include the implementation-defined clause when one hart uses an ASID of zero and another hart uses an non-zero ASID while sharing the same \newcsr{}s and VMIDs. This relaxation allows simpler hardware implementations when handling ASID 0, allowing them to simply map ASID 0 to an unique global ASID (e.g. hart ID) instead of implementing complex logic to deal with ASID 0 specially in the entire TLB hierarchy.

This design opens up space for possible designs:
\begin{itemize}
    \item If an implementation utilises global ASID space, but does not want to support LPAR, e.g. not needed or there is an hypervisor extension for the task, the implementation can simply hardwire \newcsr{} to 0.
    \item If an implementation does not exploit global ASID space, it can hardwire \newcsr{} to the hart ID (or any unique number).
    \item If an implementation wants to utilise global ASID while supporting LPAR, it will need to implement writable bits in this CSR. It also needs to add corresponding tag bits in TLB entries. Note that this is not an additional overhead as without this extension and global ASID, it would need to tag TLB entries with hart IDs anyway.
\end{itemize}

The proposed extension also allows other topologies. For example, if there are two sockets and ASID sharing is supported within a socket, then the hardwired constant bits can be set to different values on two sockets, so the software would know that some harts can share a translation cache but not others.

The proposed extension requires software to use non-zero ASIDs and global pages to represent consistent meaning across harts, unless they are aware of the \newcsr{} configuration. We looked into existing kernel ports (Linux and FreeBSD), and have seen no usage of global page and non-zero ASIDs so far, so our proposed change will not break backward compatibility of software. The proposed change will also not affect existing hardware, as accesses to the proposed CSR will trigger illegal instruction faults. The firmware could catch the exception and return the hart ID instead. Moreover, we expect that most implementations will not need to support both ASID sharing and LPAR, so they would only need to implement a readonly CSR, which is almost free in hardware.

%% file: sections/conclusion.tex
\section{Conclusion}

We present a framework for simulating different TLB organisations for RISC-V systems. Our framework can simulate multi-level TLB designs at around 400 MIPS per 8 cores, incurring a performance overhead of $\sim$20\% when compared to unmodified QEMU. We use the tool to log and categorise \sfence{}s issued by Linux and suggest an extra constraint on hardware to allow more efficient software implementation. We also investigate the case for shared TLBs and a global ASID space. We found that shared TLBs with global ASID space offer a concrete advantage over other organisations. Based on this empirical data, we propose an extension to permit such a design for RISC-V.